\documentclass[
aps,
amsmath,
amssymb,
reprint,
prl,
superscriptaddress
]{revtex4-1}

\usepackage{graphicx}
\usepackage{dcolumn}
\usepackage{bm}
\usepackage[utf8]{inputenc}
\usepackage[T1]{fontenc}
\usepackage{mathptmx}
\usepackage{etoolbox}
\usepackage{lineno}
\newcommand*\patchAmsMathEnvironmentForLineno[1]{
    \expandafter\let\csname old#1\expandafter\endcsname\csname #1\endcsname
    \expandafter\let\csname oldend#1\expandafter\endcsname\csname end#1\endcsname
    \renewenvironment{#1}
    {\linenomath\csname old#1\endcsname}
    {\csname oldend#1\endcsname\endlinenomath}}
\newcommand*\patchBothAmsMathEnvironmentsForLineno[1]{
    \patchAmsMathEnvironmentForLineno{#1}
    \patchAmsMathEnvironmentForLineno{#1*}}
\AtBeginDocument{
    \patchBothAmsMathEnvironmentsForLineno{equation}
    \patchBothAmsMathEnvironmentsForLineno{align}
    \patchBothAmsMathEnvironmentsForLineno{flalign}
    \patchBothAmsMathEnvironmentsForLineno{alignat}
    \patchBothAmsMathEnvironmentsForLineno{gather}
    \patchBothAmsMathEnvironmentsForLineno{multline}
}

\begin{document}
    \title{Magnetometry of neurons using a superconducting qubit}
    \author{Hiraku~Toida}
    \altaffiliation{Correspondence and requests for materials should be addressed to H.T.}
    \email{hiraku.toida.ds@hco.ntt.co.jp}

    \author{Koji~Sakai}

    \author{Tetsuhiko~F.~Teshima}
    \altaffiliation[Current address: ]{Medical and Health Informatics Laboratories, NTT Research Incorporated, Sunnyvale, California 94085, USA}
    \affiliation{NTT Basic Research Laboratories, NTT Corporation, 3-1 Morinosato-Wakamiya, Atsugi, Kanagawa, 243-0198, Japan}

    \author{Masahiro~Hori}
    \affiliation{Research Institute of Electronics, Shizuoka University, 3-5-1 Johoku, Naka-ku, Hamamatsu, Shizuoka 432-8011, Japan}

    \author{Kosuke~Kakuyanagi}

    \author{Imran~Mahboob}
    \affiliation{NTT Basic Research Laboratories, NTT Corporation, 3-1 Morinosato-Wakamiya, Atsugi, Kanagawa, 243-0198, Japan}

    \author{Yukinori~Ono}
    \affiliation{Research Institute of Electronics, Shizuoka University, 3-5-1 Johoku, Naka-ku, Hamamatsu, Shizuoka 432-8011, Japan}

    \author{Shiro~Saito}
    \affiliation{NTT Basic Research Laboratories, NTT Corporation, 3-1 Morinosato-Wakamiya, Atsugi, Kanagawa, 243-0198, Japan}

    \begin{abstract}
        We demonstrate magnetometry of cultured neurons on a polymeric film using a superconducting flux qubit that works as a sensitive magnetometer in a microscale area.
        The neurons are cultured in Fe$^{3+}$ rich medium to increase magnetization signal generated by the electron spins originating from the ions.
        The magnetometry is performed by insulating the qubit device from the laden neurons with the polymeric film while keeping the distance between them around several micrometers.
        By changing temperature (12.5 -- 200 mK) and a magnetic field (2.5 -- 12.5 mT), we observe a clear magnetization signal from the neurons that is well above the control magnetometry of the polymeric film itself.
        From electron spin resonance (ESR) spectrum measured at 10 K, the magnetization signal is identified to originate from electron spins of iron ions in neurons.
        This technique to detect a bio-spin system can be extended to achieve ESR spectroscopy at the single-cell level, which will give the spectroscopic fingerprint of cells.
    \end{abstract}

    \maketitle

    Iron is one of the most abundant trace elements in the human body.
    The spatial distribution and valence of iron in tissues provides us with fruitful information that can reveal the molecular and cellular mechanisms of toxicity, metabolism, and disease related to metals inside the body at the cell level \cite{Bonda2011, Jomova2011}.
    In the field of molecular biology, mass spectrometry (MS), such as inductively coupled plasma MS \cite{ISHIHARA2015} or matrix assisted laser desorption/ionization MS \cite{Schober2012}, is used for quantitative analysis of molecules at the cell level, although the requirement of cell homogenization prevents \textit{in situ} investigation.
    Optical methods such as Raman spectroscopy are frequently used as an \textit{in situ} observation tool of biomolecules, including imaging of the detailed structure of individual cell \cite{Palonpon2013}.
    Although these methods can detect metallic elements in a single cell, detailed information about metallic elements (e.g., valence change as a result of detox reactions or coupling to protein) can not be obtained.
    Another method to investigate metallic elements in cells is electron spin resonance (ESR) spectroscopy, which can resolve the oxidation state or coordination structure of ions from spectroscopic fingerprints.
    However, the limited spatial resolution and sensitivity of conventional ESR spectrometers prevent \textit{in situ} inspection at the single-cell level.

    With advances in superconducting technologies, two different types of superconducting ESR spectrometers have been actively investigated to detect a small number of electron spins either using superconducting resonators \cite{Bienfait2016, Bienfait2017, Probst2017a, Albertinale2021} or superconducting magnetometers \cite{Toida2016, Budoyo2018, Toida2019, Budoyo2020, Hirayama2021}.
    We have previously demonstrated ESR spectrometers using superconducting flux qubits which can access areas of a few micrometers with a sensitivity of  tens of spins/$\sqrt{\mathrm{Hz}}$ \cite{Budoyo2020}.

    In this work, we successfully detected the electron spins in iron-rich cultured neurons using a superconducting flux qubit.
    We attached the neurons cultured on an insulation layer to the flux qubit chip.
    As a function of temperature and an in-plane magnetic field, we observed the change in the magnetization of the cultured neurons.
    It is important to note that the spatial resolution of our sensor is close to the typical size of cells (10 -- 20 $\mu$m), which enables us to investigate the sample with cellular level resolution.
    Thus, our results tantalizingly pave the way towards realizing ESR spectroscopy at the single-cell level.
    In addition, arraying a large number of flux qubits or moving the neuron laden insulation layer with a nano-manipulator suggest the exciting prospect of imaging the distribution of spins in tissues.
    This technique will enable us to investigate metal depositions in pathological model tissues or cells for understanding cellular iron metabolism related to neuroferritinopathy, for example.

    \subsection{Results}
    \begin{figure*}
        \centering
        \includegraphics[clip]{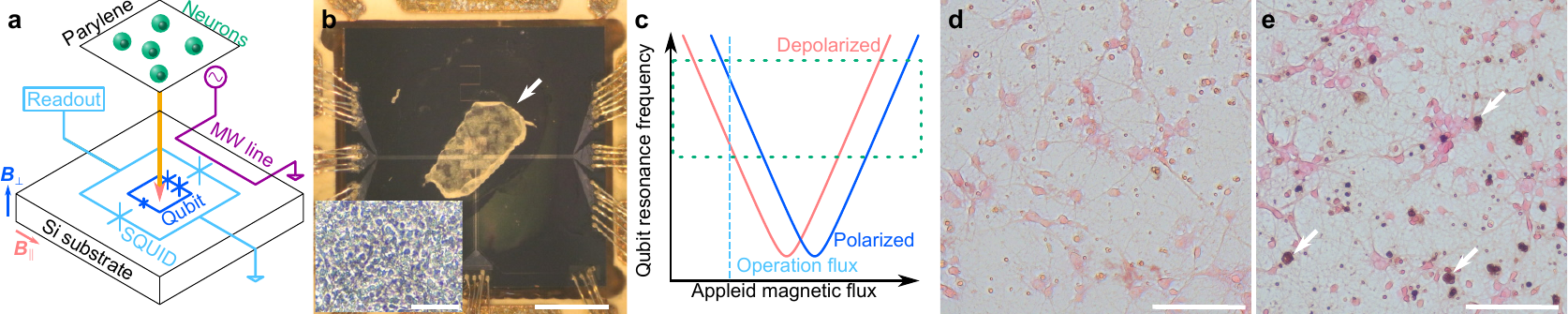}
        \caption{
        \label{fig:1}
        \textbf{Experimental setup for magnetometry of cultured neurons using a superconducting flux qubit.}
        \textbf{a} Experimental setup.
        The flux qubit (blue) works as a sensitive magnetometer.
        The quantum state of the qubit is read out by a superconducting quantum interference device (SQUID) with a room temperature readout circuit (sky blue).
        The flux qubit is excited by applying a microwave tone to an on-chip microwave (MW) line (purple).
        Cultured neurons (green) on a parylene-C film are attached to the qubit chip (orange arrow).
        An in-plane magnetic field $\bm{B}_\parallel$  (pink) is applied to polarize the electron spins in neurons, while
        a perpendicular magnetic field $\bm{B}_\perp$ (blue) is applied to control the operation flux of the qubit.
        \textbf{b} Stereomicroscope image of the qubit chip.
        The neurons cultured on a parylene-C film are indicated by the white arrow.
        Scale bar: 1 mm.
        Inset: Phase-contrast image of neurons on a parylene-C film.
        Scale bar: 100 $\mu$m.
        \textbf{c} The principle of magnetometry using a flux qubit.
        The qubit spectrum with (without) magnetic flux generated by the electron spins in neurons is shown by the blue (pink) curve.
        The magnetometry is performed by monitoring the change in the qubit resonance frequency at a fixed operation flux (sky blue).
        The green dotted rectangle corresponds to measured regions in Fig. \ref{fig:2}a.
        \textbf{d}, \textbf{e} Brightfield image of neurons cultured in medium containing \textbf{d} 0.2 $\mu$M (control) and \textbf{e} 50 $\mu$M of Fe$^{3+}$ stained by nuclear fast red (cell soma; pink) and Prussian Blue with 3,3’-diaminobenzidine (Fe$^{3+}$; dark brown).
        The white arrows in \textbf{e} indicate examples of the neurons with high Fe$^{3+}$ concentration.
        Scale bars: 100 $\mu$m.
        }
    \end{figure*}
    \subsubsection{Principle of magnetometry}
    Figure \ref{fig:1} shows our experimental setup to detect a magnetization signal from neurons.
    We use a superconducting flux qubit as a sensitive magnetometer (Fig. \ref{fig:1}a).
    On top of the qubit chip, the neurons cultured on a parylene-C film is laminated as shown in Fig. \ref{fig:1}b.
    The neurons are magnetized by applying an in-plane magnetic field $\bm{B}_\parallel$.
    This magnetization shifts the qubit spectrum by generating an additional magnetic field to the qubit (Fig. \ref{fig:1}c).
    Thus, magnetization generated by the neurons can be detected by monitoring the change in the resonance frequency of the qubit with fixed operation flux.
    In this experiment, the neurons were cultured in an Fe$^{3+}$-rich medium to enhance the signal intensity as detailed in Fig. \ref{fig:1}d and e.
    See ``Methods'' for experimental details.

    \begin{figure*}
        \centering
        \includegraphics[clip, width=170 mm]{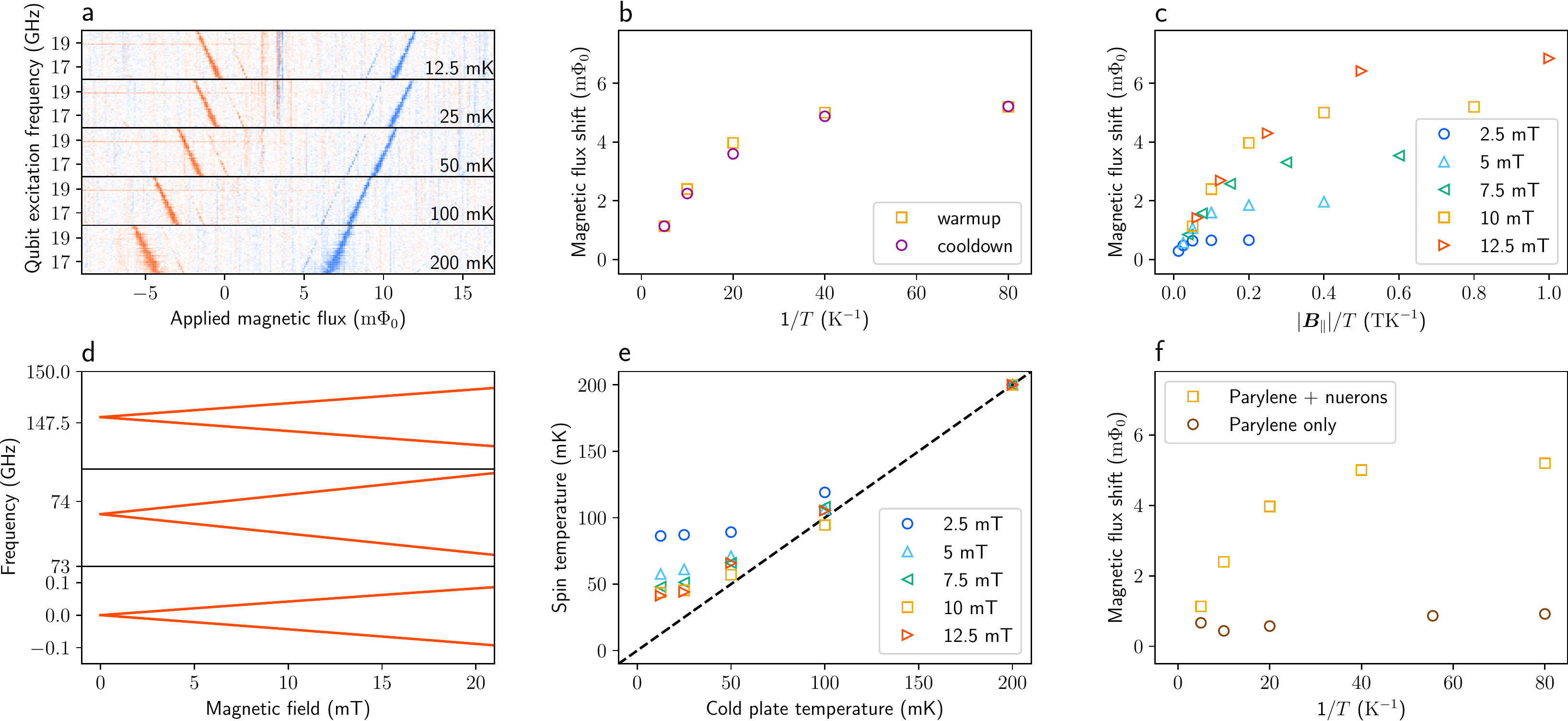}
        \caption{
        \label{fig:2}
        \textbf{Results of magnetometry.}
        \textbf{a} Spectra of the flux qubit with neurons measured at 12.5, 25, 50, 100, and 200 mK.
        The temperature was adjusted from low to high in the experiment.
        The in-plane magnetic field was fixed at 10 mT.
        The color scale indicates the switching probability of the SQUID [20\% (blue), 50\% (white), and 80\% (red)].
        The switching probability of 50\% (20 and 80\%) corresponds to the ground (excited) state of the qubit.
        \textbf{b} Temperature dependence of the magnetization with an in-plane magnetic field of 10 mT.
        The sample was first heated to 200 mK from base temperature (red squares).
        The sample was then cooled from 200 mK to verify the experimental results (purple circles).
        \textbf{c} Temperature and in-plane magnetic field dependence of the magnetization as a function of $|\bm{B}_\parallel|/T$.
        \textbf{d} Calculated energy diagram of electron spins in neurons as a function of magnetic field.
        The direction of the magnetic field was $+z$ for this calculation.
        \textbf{e} Effective spin temperature as a function of the cold plate temperature.
        The dashed line is a guide for the eye, which indicates that the spin temperature is equal to the cold plate temperature.
        \textbf{f} Comparison of magnetization signal of parylene-C film with neurons and pure parylene-C film under an in-plane magnetic field of 10 mT.
        Error bars in \textbf{b}, \textbf{c}, \textbf{e}, and \textbf{f} are smaller than the symbols.
        }
    \end{figure*}
    \subsubsection{Magnetometry of neurons}
    Figure \ref{fig:2} summarizes the measurement results of magnetization from neurons.
    The magnetometry was performed at different temperatures $T$ and in-plane magnetic fields $|\bm{B}_\parallel|$ to control the polarization of the electron spins in the neurons.
    The qubit spectra as a function of temperature at fixed in-plane magnetic field of 10 mT are shown in Fig. \ref{fig:2}a.
    Due to parasitic circuit resonances, we only observed a clear qubit spectrum above 16 GHz.
    Overall, the spectra shift to the negative flux side with increasing temperature.
    This shift suggests that the qubit senses the decrease of the magnetization from the electron spins in the neurons with increasing temperature.

    The amount of the flux shift as a function of the inverse temperature is summarized in Fig. \ref{fig:2}b.
    In the high-temperature range, the magnetization, namely the magnetic flux shift increases almost linearly, while in the low-temperature range, the magnetization saturates.
    Figure \ref{fig:2}b also indicates that the magnetization curve is well reproduced in both temperature sweep directions (warmup and cooldown) without hysteresis.
    It is important to note that a waiting time of more than 45 minutes is necessary after changing the temperature to thermalize the sample.

    To investigate the temperature dependence in more detail, we measured the in-plane magnetic field dependence of the magnetization as shown in Fig. \ref{fig:2}c.
    Here, the magnetization is plotted as a function of the ratio $|\bm{B}_\parallel|/T$.
    In general, the magnetization is determined by temperature and the magnetic field.
    For example, the simplest two-level spin system with Zeeman energy which linearly depends on a magnetic field $B$ has hyperbolic-tangent-dependent magnetization as a function of the ratio $B/T$.
    However, electron spins induced by Fe$^{3+}$ ions in cells are reported to have a more complex energy level structure expressed by the Hamiltonian \cite{Kumar2016}
    \begin{align}
        H=\mu_B\bm{B}g\bm{S}+DS_z^2+E(S_x^2-S_y^2),
    \end{align}
    where $\mu_B$ is the Bohr magneton, $\bm{B} = \bm{B}_\perp + \bm{B}_\parallel $ is the magnetic field,  $g$ is the anisotropic g-factor tensor, $\bm{S}=(S_x, S_y, S_z)$ is the spin operator with the spin angular momentum $S$, $D$ is the zero-field splitting, and $E$ is the axial splitting.
    We used spin angular momentum $S=5/2$, and the values of $g$, $D$, and $E$ reported in the literature \cite{Kumar2016} to calculate the energy spectrum.
    It should be noted that the alignment between the in-plane magnetic field and the principal axis of the spin Hamiltonian is random because the neurons are randomly cultured on the parylene-C film.
    Thus, the magnetization curves should be analyzed considering both the complex energy level structure and the randomness of the alignment.

    An example of the energy level structure is plotted in Fig. \ref{fig:2}d.
    Since the Hamiltonian has $S=5/2$, it will result in six Zeeman-split levels.
    However, the system can be approximated by a two-level system in the magnetic field and temperature range for this magnetometry experiment.
    This is due to the large energy splitting of more than 70 GHz between the first and second excited states, which is much larger than the energy scale of this experiment.
    As a result, electrons only occupy the ground and first excited states.
    To derive the magnetization from numerical simulations, we averaged the contribution from all solid angles between the magnetic field and the principal axis of the spin.
    The simulation shows that the magnetization increases as a function of the ratio $B/T$ (Supplementary Fig. 2) in our experimental conditions.
    This means the data for different magnetic fields and temperatures is explained by one curve.

    However, the observed magnetization shown in Fig. \ref{fig:2}c shows a deviation from the aforementioned simulation which indicates linear increase in magnetization.
    We attribute this variation to the temperature of the spins in neurons deviating from the measured temperature of the cold plate in the dilution refrigerator.
    This might originate from the relatively low thermal conductivity of a parylene-C film, whose room-temperature value is 0.082 Wm$^{-1}$K$^{-1}$.
    Correcting for this deviation by assuming that there is no temperature discrepancy between the spin and thermometer temperature at 200 mK, we derive the effective spin temperature as shown in Fig. \ref{fig:2}e.
    This reveals the spin temperature saturates around 40 -- 90 mK in the low-temperature range.
    The saturation temperatures are different for the different magnetic fields.
    This observation might be related to the magnetic-field-dependent spin-lattice relaxation rate, as observed in solid state materials \cite{Davids1964}, but a detailed discussion is beyond the scope of this letter.
    It should be noted that magnetometry of paramagnetic material is used as a thermometer in low-temperature ranges \cite{Hirschkoff1970}.

    \subsubsection{Magnetometry of insulator film}
    To confirm the magnetization signal originates from the neurons, we performed magnetometry of pure parylene-C film.
    Parylene-C films are reported to have unpaired electron spins with g-factor $g^{\prime}\sim$ 2 when they are damaged by an electron beam \cite{Senkevich2004}.
    Figure \ref{fig:2}f compares the results of magnetometry of the parylene-C film containing neurons with those of pure parylene-C film.
    This clearly shows that the magnetization signal from the parylene-C film is much smaller than the one originating from the neurons.
    Thus, we conclude that the magnetometry shown in Figs. \ref{fig:2}a -- c does indeed originate from the electron spins in the cultured neurons.

    \begin{figure}
        \centering
        \includegraphics[clip, width=0.4\textwidth]{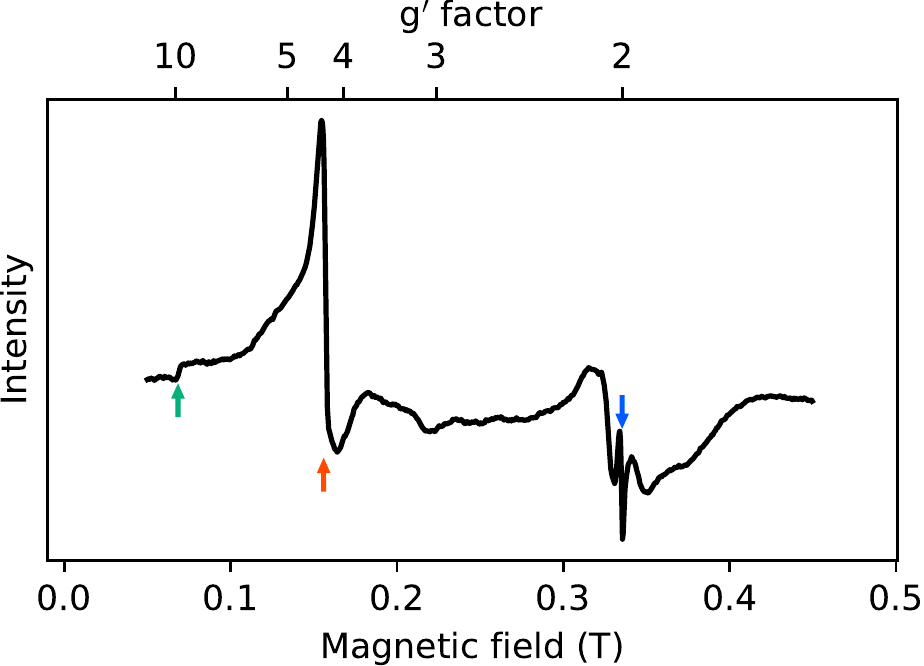}
        \caption{
            \label{fig:3}
            \textbf{Result of ESR spectroscopy of neurons measured with a conventional X-band ESR spectrometer.}
            The temperature of the sample is fixed at 10 K.
            Green, red, and blue arrow indicate ESR signals corresponding to $g^{\prime}=$ 9.8, 4.3, and 2.0, respectively.
            It should be noted that the shape of the spectrum is derivative due to the modulated magnetic field.
        }
    \end{figure}
    \subsubsection{Conventional ESR spectroscopy of neurons}
    In addition, conventional ESR spectroscopy was performed for neurons cultured in the same condition as the qubit experiments to verify the origin of the magnetization signal from the neurons.
    The spectrum shown in Fig. \ref{fig:3} has three large peaks corresponding to $g^\prime=$ 9.8, 4.3, and 2.0.
    The largest peak with $g^\prime=$ 4.3 is the signature of high spin state iron in cells with $\pm 3/2$ transition, while the small peak with $g^\prime=$ 9.8 can originate from $\pm 1/2$ transition \cite{Kumar2016}.
    The peak with $g^\prime=$ 2.0 might be interpreted as the peak of other metal ions, mostly copper, and radicals \cite{Kumar2016}.
    Considering the peak height of the spectra, the contribution to the magnetization signal mainly comes from iron ions.
    This ESR spectrum thus conclusively confirms that the magnetization signal of iron ions in neurons can be measured by a superconducting flux qubit.

    \subsection{Discussion}
    To estimate the number of measured neurons, we calculated the average distance between neurons to be 14 $\mu$m from the seeding condition (Fig.\ref{fig:1}b; inset).
    This value is similar to the qubit size of 24$\times$6 $\mu$m.
    Comparison of these two values leads to the conclusion that the sensing area of the qubit is expected to only have one neuron.
    Thus, this method of magnetization detection is applicable to investigating the properties of a single neuron.

    In general, it is possible to obtain material parameters from a magnetization curve, e.g., the spin angular momentum or g-factor.
    These parameters are key to identifying spin species in a sample.
    To obtain them, information about the magnetization curve, including the saturation regime, should be measured by lowering the temperature or applying a larger in-plane magnetic field.
    As discussed earlier, the spin temperature saturates at more than 40 mK, which is higher than the cold plate temperature.
    According to the simulated magnetization curve (Supplementary Fig. 2), we can observe the deviation from the linear dependence of the magnetization if the spins are cooled down to 12.5 mK at the magnetic field of 12.5 mT as indicated by the arrow in Supplementary Fig. 2.
    Consequently, using materials with higher thermal conductivity for insulation offers a route towards deriving the material parameters.
    The simulation also implies that the condition of 12.5 mK and 12.5 mT is still inadequate for observing the full saturation of the magnetization.
    Thus, application of a larger in-plane magnetic field is necessary to reach full saturation.
    However, the flux qubit in this work is made of aluminum, which becomes inoperational for in-plane fields greater than 20 mT.
    To provide more insight into bio-spin systems, future development of flux qubits with compatibility to higher magnetic fields is required by, for instance, using niobium-based materials or thin superconducting film \cite{Krause2022}.

    In addition, the magnetization curve should give a number of electron spins in the sensing volume, as we demonstrated in previous experiments \cite{Toida2016, Budoyo2018, Toida2019, Budoyo2020, Hirayama2021}.
    However, it was difficult to quantify the electron spins in neurons exactly in the present experiment.
    This is due to the lack of conversion factor between magnetization and the number of spins.
    Such a conversion factor is attainable if the magnetometry up to the saturation regime is performed with a reference sample whose concentration is already known.
    This kind of quantitative analysis is especially important for expanding this method to spin imaging.
    For example, by comparing the results of magnetometry with stained microscope images like those in Figs.\ref{fig:1}d and e, we can compare the distribution of electron spins whose parameters (e.g., g-factor) are identified with that of iron ions.

    Although an exact conversion factor from magnetization to the number of spins is unknown,
    the number of detected spins can still be estimated by assuming the conversion factor from solid state materials \cite{Toida2016, Budoyo2018, Toida2019, Budoyo2020, Hirayama2021}.
    This analysis yields the number of detected spins in neurons as 9$\times$10$^6$.
    The detection volume for the qubit magnetization measurements is approximately 100 $\mu$m$^3$ and hence a spin density of 9$\times10^{13}$ spins/mm$^3$ can be extracted.
    In addition, the conventional ESR spectroscopy utilized in Fig.\ref{fig:3} can also render an estimate for the number of spins per unit volume of 2$\times10^{12}$ -- 2$\times10^{13}$ spins/mm$^3$.
    Consequently these two very different measurement protocols yield similar order of magnitude spin densities thus indicating their common origin.
    It should be noted that the larger spin density acquired from the ESR spectroscopy is due to the looser packing of cells in the large ESR tube.

    The experimentally determined spin density can also be used to calculate the mass of iron in unit weight of cells in the neuron sample yielding 8 $\mu$g/g.
    This result can be compared to ESR spectroscopy of human brain tissue from literature which indicates an iron mass weight in the range 2 -- 34 $\mu$g/g \cite{Kumar2016, Vroegindeweij2021}.
    Consequently the overlap in these numbers confirms that the flux qubit does indeed detect magnetization from iron ions that is located in neurons.

    In conclusion, we have demonstrated the detection of electron spins in cultured neurons using a superconducting flux qubit.
    By combining the ESR spectra obtained with a conventional ESR spectrometer, the spin species is identified as iron in neurons.
    We emphasize that the typical size of the cells is close to the loop size of the qubit, which enables the investigation of individual neurons.
    Our device is also capable of detecting ESR signals if the excitation tone for the neurons is applied, as we previously demonstrated for solid-state spin materials \cite{Toida2016, Budoyo2018, Toida2019, Budoyo2020, Hirayama2021}.
    Thus, the more detailed characteristics of the neurons would be visible from spectroscopic fingerprints in future experiments, including the detection of other metallic elements such as chromium, manganese, or copper.

    \subsection{Acknowledgments}
    We thank Y.~Furukawa for assistance with the cell culture.
    This work was supported by CREST (Grant No. JPMJCR1774).

    \subsection{Methods}
    \subsubsection{Experimental setup}
    Figure \ref{fig:1}a shows the experimental setup for magnetometry of cultured neurons.
    A superconducting flux qubit was fabricated on a silicon substrate.
    To read out the quantum state of the qubit, we inductively coupled a superconducting quantum interference device (SQUID) to the qubit \cite{Deppe2007}.
    The qubit was excited by a microwave tone applied through an on-chip microwave line.
    We attached the neurons cultured on a polymeric film to the qubit chip as shown in Fig. \ref{fig:1}b.

    Flux qubits are two-level systems with a hyperbolic resonance frequency dependence as a function of the perpendicular magnetic field through the loop of the qubit $\bm{B}_\perp$ \cite{Mooij1999, Orlando1999} (Fig. \ref{fig:1}c).
    To convert the change in the magnetization of the neurons into the change in the resonance frequency of the qubit, the qubit spectrum should have a finite slope.
    This condition is satisfied by adjusting the operation flux of the qubit as shown in Fig. \ref{fig:1}c.
    Details of the methodology for solid-state spin materials are available in our previous papers \cite{Toida2016, Budoyo2018, Toida2019, Budoyo2020, Hirayama2021}.

    The major difference between solid-state spin materials and neurons is the necessity of insulation between the spin sample and qubit device.
    We first confirmed that the resistance of the SQUID shows an irregular response at low temperatures for a device with neurons if no insulation layer is inserted between them.
    Thus, the measurement of the qubit state can not be performed without insulation.

    To insulate the qubit device from neurons, a poly(chloro-\textit{p}-xylylene) (parylene-C) film with the thickness of 2 $\mu$m is laminated on the qubit surface.
    Parylene-C is the mobile substrates for adherent cells with high biocompatibility, stiffness, optical transparency, and high electrical resistance \cite{Teshima2014, Teshima2016}.
    These features are ideal to transfer the film with cells onto the qubit chip while keeping electrical insulation between them.
    Rat hippocampal neurons are cultured on a parylene-C film and then fixed with glutaraldehyde by cross-linking the proteins.
    The neuron-laden parylene-C film is transferred onto the qubit chip.
    To make the alignment of neurons to the qubit easy, neurons are densely seeded on the parylene-C film with an initial density of 5000 cells/mm$^2$ (Fig. \ref{fig:1}b; inset).

    For this experiment, we cultured the neurons in an Fe$^{3+}$-rich culture medium to increase the number of the intracellular Fe$^{3+}$ ions, as shown in Figs. \ref{fig:1}d and e.
    As indicated by the white arrows in Fig. \ref{fig:1}e, some neurons uptake a high amount of Fe$^{3+}$ ions.
    The elevated level of intracellular Fe$^{3+}$ (Supplementary Fig. 1a) and the increased number of Fe$^{3+}$-rich neurons (Supplementary Fig. 1b)  indicate that the culture in the Fe$^{3+}$-rich medium results in high intracellular Fe$^{3+}$ ions.

    The flux qubit works as a magnetometer for unpaired electron spins originating from Fe$^{3+}$ ions.
    To enhance the sensitivity of the magnetometry, the distance between the qubit and neurons should be minimized.
    To this end, after transferring the neuron-laden parylene-C film on the qubit chip, a small amount of ethanol was introduced to induce attachment via surface tension.
    Judging from an optical interference pattern formed by the gap between the qubit chip and the parylene-C film \cite{Zhu2011}, the typical distance between the qubit and neurons is estimated to be several micrometers, which is within the sensing volume of the flux qubit.

    \subsection{Data availability}
    The data that support the plots in this paper are available from the corresponding
    author on reasonable request.

    \subsection{Author contributions}
    All the authors contributed extensively to the work presented in this paper.
    H.T., K.S., T.T. and I.M. conceptualized the experiment.
    H.T. and K.S. carried out the measurements and data analysis.
    K.K., H.T and S.S. designed and developed the flux-qubit-based magnetometry system.
    K.S. and T.T. prepared the neurons cultured on parylene films.
    M.H. and Y.O. performed ESR spectroscopy.
    H.T. and K.S. wrote the paper, with feedback from all the authors.
    S.S. supervised the project.

    \subsection{Competing interests}
    The authors declare no competing interests.


\begin{thebibliography}{100}

    \bibitem{Bonda2011}
    Bonda, D.~J. et~al. {Role of metal dyshomeostasis in Alzheimer's disease†}.
    {\em Metallomics}{ \bf 3}, 267-270 (2011).

    \bibitem{Jomova2011}
    Jomova, K. \& Valko, M. Advances in metal-induced oxidative stress and human
    disease. {\em Toxicology}{ \bf 283}, 65-87 (2011).

    \bibitem{ISHIHARA2015}
    Ishihara, Y. et~al. Development of desolvation system for single-cell analysis
    using droplet injection inductively coupled plasma atomic emission
    spectroscopy. {\em Analytical Sciences}{ \bf 31}, 781-785 (2015).

    \bibitem{Schober2012}
    Schober, Y., Guenther, S., Spengler, B. \& Römpp, A. Single cell
    matrix-assisted laser desorption/ionization mass spectrometry imaging. {\em
        Anal. Chem.}{ \bf 84}, 6293--6297 (2012).

    \bibitem{Palonpon2013}
    Palonpon, A.~F. et~al. Raman and {SERS} microscopy for molecular imaging of
    live cells. {\em Nature Protocols}{ \bf 8}, 677--692 (2013).

    \bibitem{Bienfait2016}
    Bienfait, A. et~al. Reaching the quantum limit of sensitivity in electron spin
    resonance. {\em Nat. Nanotechnol.}{ \bf 11}, 253--257 (2016).

    \bibitem{Bienfait2017}
    Bienfait, A. et~al. Magnetic resonance with squeezed microwaves. {\em Phys.
        Rev. X}{ \bf 7}, 041011 (2017).

    \bibitem{Probst2017a}
    Probst, S. et~al. Inductive-detection electron-spin resonance spectroscopy with
    65 spins/{H}z sensitivity. {\em Appl. Phys. Lett.}{ \bf 111}, 202604 (2017).

    \bibitem{Albertinale2021}
    Albertinale, E. et~al. Detecting spins by their fluorescence with a microwave
    photon counter. {\em Nature}{ \bf 600}, 434--438 (2021).

    \bibitem{Toida2016}
    Toida, H. et~al. Electron paramagnetic resonance spectroscopy using a direct
    current-{SQUID} magnetometer directly coupled to an electron spin ensemble.
    {\em Appl. Phys. Lett.}{ \bf 108}, 052601 (2016).

    \bibitem{Budoyo2018}
    Budoyo, R.~P. et~al. Electron paramagnetic resonance spectroscopy of
    {E}r$^{3+}$:{Y}$_{2}${S}i{O}$_{5}$ using a {J}osephson bifurcation amplifier:
    Observation of hyperfine and quadrupole structures. {\em Phys. Rev. Mater.}{
        \bf 2}, 011403 (2018).

    \bibitem{Toida2019}
    Toida, H. et~al. Electron paramagnetic resonance spectroscopy using a single
    artificial atom. {\em Commun. Phys.}{ \bf 2}, 33 (2019).

    \bibitem{Budoyo2020}
    Budoyo, R.~P., Kakuyanagi, K., Toida, H., Matsuzaki, Y. \& Saito, S. Electron
    spin resonance with up to 20 spin sensitivity measured using a
    superconducting flux qubit. {\em Appl. Phys. Lett.}{ \bf 116}, 194001 (2020).

    \bibitem{Hirayama2021}
    Hirayama, Y., Ishibashi, K. \& Nemoto, K., editors.
    \newblock {\em Hybrid Quantum Systems}.
    \newblock Springer, Singapore,  (2021).

    \bibitem{Kumar2016}
    Kumar, P. et~al. A novel approach to quantify different iron forms in ex-vivo
    human brain tissue. {\em Sci. Rep.}{ \bf 6}, 38916 (2016).

    \bibitem{Davids1964}
    Davids, D.~A. \& Wagner, P.~E. Magnetic field dependence of paramagnetic
    relaxation in a {K}ramers salt. {\em Phys. Rev. Lett.}{ \bf 12}, 141--142
    (1964).

    \bibitem{Hirschkoff1970}
    Hirschkoff, E.~C., Symko, O.~G., Vant-Hull, L.~L. \& Wheatley, J.~C.
    Observation of the static nuclear magnetism of pure metallic copper in low
    magnetic fields. {\em Journal of Low Temperature Physics}{ \bf 2}, 653--665
    (1970).

    \bibitem{Senkevich2004}
    Senkevich, J.~J. et~al. Correlation between bond cleavage in parylene {N} and
    the degradation of its dielectric properties. {\em Electrochem. Solid-State
        Lett.}{ \bf 7}, G56-G58 (2004).

    \bibitem{Krause2022}
    Krause, J. et~al. Magnetic field resilience of three-dimensional transmons with
    thin-film {Al/AlO}$_{x}$/{Al} {J}osephson junctions approaching 1 {T}. {\em
        Phys. Rev. Applied}{ \bf 17}, 034032 (2022).

    \bibitem{Vroegindeweij2021}
    Vroegindeweij, L.~H. et~al. Quantification of different iron forms in the
    aceruloplasminemia brain to explore iron-related neurodegeneration. {\em
        NeuroImage: Clinical}{ \bf 30}, 102657 (2021).

    \bibitem{Deppe2007}
    Deppe, F. et~al. Phase coherent dynamics of a superconducting flux qubit with
    capacitive bias readout. {\em Phys. Rev. B}{ \bf 76}, 214503 (2007).

    \bibitem{Mooij1999}
    Mooij, J.~E. et~al. Josephson persistent-current qubit. {\em Science}{ \bf
        285}, 1036--1039 (1999).

    \bibitem{Orlando1999}
    Orlando, T.~P. et~al. Superconducting persistent-current qubit. {\em Phys. Rev.
        B}{ \bf 60}, 15398--15413 (1999).

    \bibitem{Teshima2014}
    Teshima, T. et~al. Parylene mobile microplates integrated with an enzymatic
    release for handling of single adherent cells. {\em Small}{ \bf 10}, 912-921
    (2014).

    \bibitem{Teshima2016}
    Teshima, T. et~al. High-resolution vertical observation of intracellular
    structure using magnetically responsive microplates. {\em Small}{ \bf 12},
    3366-3373 (2016).

    \bibitem{Zhu2011}
    Zhu, X. et~al. Coherent coupling of a superconducting flux qubit to an electron
    spin ensemble in diamond. {\em Nature}{ \bf 478}, 221--224 (2011).

\end{thebibliography}
\end{document}